\documentclass{article}
\usepackage{spconf,amsmath,graphicx,hyperref}
\usepackage{amssymb}
\usepackage{booktabs}
\usepackage{multirow}

\usepackage{xcolor}

\title{MT-HuBERT: Self-Supervised Mix-Training for Few-Shot Keyword Spotting in Mixed Speech}
\name{
  Junming Yuan$^{1,2}$,
  Ying Shi$^{3,2}$,
  Dong Wang$^{2}$,
  Lantian Li$^{4}$,
  Askar Hamdulla$^{1}$
}
\address{
  $^{1}$ School of Computer Science and Technology, Xinjiang University, China \\
  $^{2}$ Center for Speech and Language Technologies, BNRist, Tsinghua University, China \\
  $^{3}$ School of Computer Science and Technology, Harbin Institute of Technology, China \\
  $^{4}$ School of Artificial Intelligence, Beijing University of Posts and Telecommunications, China \\
}
\begin{document}
\ninept
\maketitle
\begin{abstract}
Few-shot keyword spotting aims to detect previously unseen keywords with very limited labeled samples. A pre-training and adaptation paradigm is typically adopted for this task. While effective in clean conditions, most existing approaches struggle with mixed keyword spotting---detecting multiple overlapping keywords within a single utterance---a capability essential for real-world applications. We have previously proposed a pre-training approach based on Mix-Training (MT) to tackle the mixed keyword detection problem and demonstrated its efficiency. However, this approach is fully supervised, unable to utilize vast unlabeled data. To this end, we propose \textbf{Mix-Training HuBERT (MT-HuBERT)}, a self-supervised learning (SSL) pre-training framework that implements the MT criterion during pre-training.
MT-HuBERT predicts, in a self-supervised manner, the clean acoustic units of each constituent signal from contextual cues, in contrast to predicting compositional patterns of mixed speech. Experiments conducted on the Google Speech Commands (GSC v2) corpus demonstrate that our proposed MT-HuBERT consistently outperforms several state-of-the-art baselines in few-shot KWS tasks under both mixed and clean conditions.
\end{abstract}
\section{Introduction}
\label{sec:intro}

Few-shot keyword spotting (KWS) addresses detection with limited labeled samples and has witnessed significant advances in recent years~\cite{lopez2021deep,tabibian2020survey}. One promising approach employs a \emph{pre-training and adaptation} framework~\cite{jung2023metric,mazumder2021few,seo2021wav2kws,lin2020training,rusci2023few}, which leverages a large amount of data to learn general speech representations and then adapts the KWS model using limited examples (typically 5, 10, or 15) per user-defined keyword.

Despite its simplicity and efficiency, most benchmark evaluations are conducted on controlled, idealized test sets that differ from real-world usage. This gap is pronounced when a KWS model is confronted with detecting multiple overlapping keywords simultaneously (e.g., ``help'' and ``call the police'' within one detection window), a hallmark of \textbf{mixed-speech} scenarios. Mix-Training (MT)~\cite{shi2023spot} offers a promising strategy that enables simultaneous detection of two or more keywords, regardless of their relative energy levels. Our previous work~\cite{yuan2024few} demonstrated the potential of MT for few-shot KWS under mixed conditions; specifically, combining SSL backbones (wav2vec~2.0~\cite{baevski2020wav2vec}, HuBERT~\cite{hsu2021hubert}, both pre-trained on \emph{single-speaker, clean} speech) with a lightweight classifier adapted by MT achieves competitive performance in both clean and 2-mix scenarios. These findings motivate integrating MT directly into SSL pre-training.

In this paper, we propose a self-supervised pre-training framework for mixed speech: Mix-Training HuBERT (MT-HuBERT). This approach enables the SSL model to disentangle and capture the clean acoustic units of each constituent signal in the mixture rather than memorizing the compositional patterns of mixed speech. We hypothesise that representing mixed speech by composition of clean units rather than compositional mixed patterns leads to better generalization. In fact, this is how our auditory system works: we always try to extract clean patterns rather than mixed patterns in a complex speech environment. 

The training process of MT-HuBERT consists of two phases: (1) for an $n$-source mixture, we construct per-frame $n$-hot targets by tokenizing each clean source with the same codebook derived from clean speech via k-means; (2) a masked-prediction scheme with a multi-label binary cross-entropy objective encourages the model to predict the active units of all the sources at the masked frame. In this way, the backbone is explicitly trained to represent mixed speech using composition of clean acoustic units rather than mixed patterns. After the pre-training, we freeze the MT-HuBERT backbone and attach two linear layers for few-shot KWS adaptation. Experiments on the GSC v2 dataset show that MT-HuBERT achieves the best performance on both clean and 2-mix data compared with several state-of-the-art SSL models, and it exhibits strong generalizability to challenging 3-mix scenarios, demonstrating the effectiveness of this self-supervised mix-training framework.

\begin{figure*}[!hbt]
  \centering
  \vspace{-6mm}
  \includegraphics[width=0.78\linewidth]{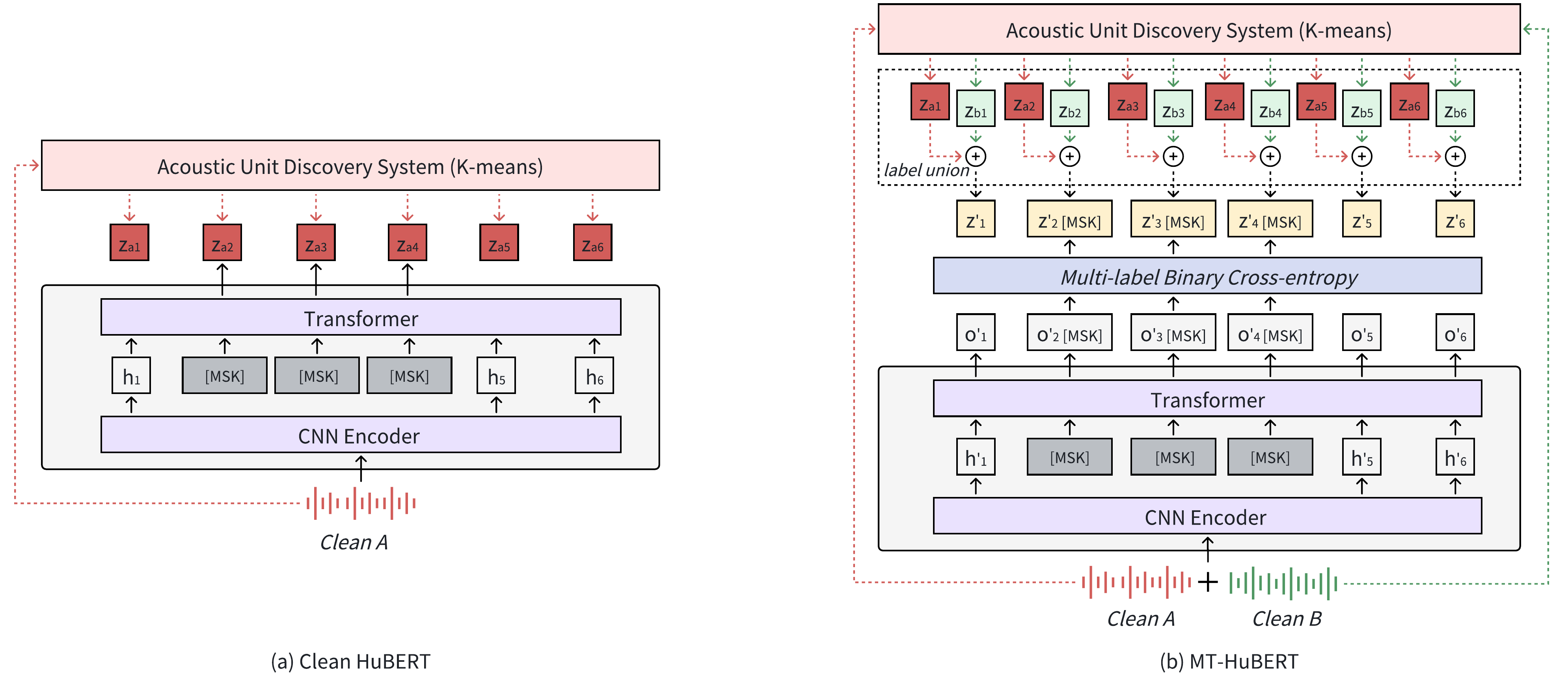}
  \vspace{-2mm}
  \caption{Comparison between clean HuBERT and the proposed MT-HuBERT.}
  \label{fig:Fig1}
  \vspace{-2mm}
\end{figure*}

\section{Related Work}

Self-supervised learning (SSL) models have proven effective across a range of speech tasks, including automatic speech recognition~\cite{baevski2020wav2vec,hsu2021hubert,zanon2024mhubert,yoon2024hubert,shankar2025selective}, emotion recognition~\cite{amiriparian2024exhubert,gong2025learning,sampath2025efficient}, speaker verification~\cite{wang2021fine,lepage2025ssps}, and speech separation~\cite{huang2022investigating}. Conventional SSL models are typically pre-trained on single-speaker, clean speech. More recently, mixture-aware SSL has attracted increasing attention.

For instance, WavLM~\cite{chen2022wavlm} creates artificial mixtures by combining utterances from different speakers and adopts a denoising-style objective that emphasizes a dominant speaker (e.g., the longest-duration one), encouraging the model to treat non-dominant components (including other speakers) as interference during masked prediction. This approach cannot present acoustic units of non-dominant sources and so does not address the mixing problem. Mixture Predictive Coding (MPC)~\cite{wang2024speech} performs pre-training as usual but on mixed speech. This means that the pre-training task is to predict mixed patterns rather than clean acoustic units. Cocktail HuBERT~\cite{fazel2023cocktail} employs permutation-invariant training (PIT)~\cite{yu2017permutation,yu2017recognizing} with a multi-head HuBERT architecture, where each branch predicts the clean acoustic units of one source. Although PIT resolves label permutation, multi-head training introduces inter-branch competition and additional architectural complexity; in practice, predicting one unit on a branch decreases the probability of other units that appear and being predicted on other branches, which can complicate optimization and calibration.


Compared with the above, our proposed MT-HuBERT performs single-head, multi-label ($k$-hot) masked prediction against a clean-speech codebook: it directly predicts \emph{which clean units are present} at each masked frame of the mixture. This avoids memorizing compositional patterns (cf.\ MPC) and sidesteps permutation/competition issues and overhead associated with multi-head PIT (cf.\ Cocktail HuBERT), providing a simpler and theoretically sound route to source-wise unit disentanglement in mixed speech.



\section{Methodology}
\label{sec:Method}

\subsection{Revisit Mix-Training}
\label{ssec:MT}

Mix-Training (MT)~\cite{shi2023spot} comprises three components: \emph{uniform mixing}, \emph{label union}, and \emph{binary cross-entropy (BCE)} training. Uniform mixing approximates real acoustic overlaps with variable energy ratios; label union uses an $n$-hot scheme to represent the coexistence of multiple sources in a mixture. Take 2-mix as an example, its data construction in MT is:
\begin{equation}
  \begin{split}
  x_{\text{mix}} &= \omega_{1} x_{i} + \omega_{2} x_{j},\\
  y_{\text{mix}} &= y_{i} \oplus y_{j},
  \end{split}
 \label{eq:mt}
\end{equation}
\noindent where $\omega_1$ and $\omega_2$ are independent scale factors and $\oplus$ denotes logical OR (label union) over one-hot label vectors. In MT, $\omega_1,\omega_2 \sim \mathcal{U}(0.1,\,0.9)$, allowing broad yet audible energy variations while avoiding vanishing components. In practice, clean speech is also interleaved during training to ensure exposure to non-mixed (i.e., clean) inputs. Models are trained in a \emph{multi-label} (sigmoid) setting with BCE, where each output dimension estimates the presence probability of a source component (e.g., a keyword).

\subsection{MT-HuBERT}
\label{ssec:MT-HuBERT}

\subsubsection{Clean HuBERT}

First, we briefly review HuBERT~\cite{hsu2021hubert} (Fig.~\ref{fig:Fig1}a). HuBERT consists of a local feature extractor $f(\cdot)$ (CNN Encoder) and a context network $g(\cdot)$ (Transformer). Given frame-level input $X=[x_1,\ldots,x_T]$ and discrete targets $Z=[z_1,\ldots,z_T]$ obtained via \emph{k}-means clustering, let $H=f(X)=[h_1,\ldots,h_T]$ be local features. A random masking operator $MSK(\cdot)$ produces $H_m=MSK(H)=[h_{m1},\ldots,h_{mT}]$, which is fed into the context network to yield $O=g(H_m)=[o_1,\ldots,o_T]$. HuBERT predicts the targets on the \emph{masked} frames by minimizing the negative log-likelihood:
\begin{equation}
     \begin{split}
     L_m = -\sum_{t\in MSK} \log p(z_t \mid o_t),
     \end{split}
     \label{eq: hubert lm}
\end{equation}
\noindent where the posterior over the codebook $\{1,\ldots,C\}$ is
\begin{equation}
  \begin{split}
   p(z_t{=}c \mid o_t) = \frac{\exp\!\big(\mathrm{sim}(A o_t, e_{c})/\tau\big)}{\sum_{c'=1}^{C} \exp\!\big(\mathrm{sim}(A o_t, e_{c'})/\tau\big)},
  \end{split}
 \label{eq: hubert p_t}
\end{equation}
\noindent with projection matrix $A$, codebook embeddings $e_c$, cosine similarity $\mathrm{sim}(\cdot,\cdot)$, and temperature $\tau$. (By abuse of notation, $MSK$ also denotes the index set of masked positions.)

\subsubsection{Proposed MT-HuBERT}
MT-HuBERT (Fig.~\ref{fig:Fig1}b) injects the MT principle into SSL pre-training. As in HuBERT, a clean-speech codebook is first derived by k-means; mixed speech with $k$ sources is then explained via $n$-hot supervision at masked frames.

Let $X'=[x'_1,\ldots,x'_T]$ be a $n$-source mixture and $Z'=[z'_1,\ldots,z'_T]$ the corresponding $n$-hot target sequence over the same codebook, where $z'_{t}\in\{0,1\}^{C}$ indicates all discrete units present at frame $t$ (label union of per-source tokenizations). The local features $H'=f(X')$ are masked to $H'_m=MSK(H')$, and $O'=g(H'_m)$ are contextual outputs. MT-HuBERT replaces the single-label softmax with a multi-label (sigmoid) classifier over units and minimizes the BCE on masked positions:
\begin{equation}
  \begin{split}
  L_m = -\sum_{t\in MSK}\sum_{c=1}^{C}\Big[ z'_{t,c}\log p_{t,c} + (1 - z'_{t,c})\log(1 - p_{t,c}) \Big],
 \end{split}
 \label{eq: mt-hubert lm}
\end{equation}
\noindent where the per-unit probability is
\begin{equation}
  \begin{split}
  p_{t,c} = \sigma(\mathrm{sim}(A' o'_t, e_{c})/\tau\big),
  \end{split}
\label{eq: mt-hubert p_t}
\end{equation}
\noindent $\sigma(\cdot)$ is the logistic sigmoid, $A'$ is a projection matrix, and $\tau$ is set to 0.1 in our experiments. This objective encourages the model to \textbf{recover all active units at a masked frame from context}, rather than memorize mixture-specific configurations. When $n=1$, MT-HuBERT reduces to standard HuBERT training on clean speech.

\section{Experiment Settings}
\label{sec:experiments}

\subsection{Datasets}
\label{ssec:Datasets}

\textbf{Pre-training dataset.} We pre-trained MT-HuBERT on the LibriSpeech-960h corpus~\cite{panayotov2015librispeech}. Data mixing followed the MT scheme and mixtures were limited to 2-mix; clean utterances were interleaved to preserve coverage of non-mixed inputs.

\noindent \textbf{Fine-tuning dataset.} Few-shot KWS evaluation was conducted on the Google Speech Commands (GSC v2) dataset~\cite{warden2018speech}. The preparation mirrored pre-training: only clean and 2-mix data were used. We fine-tuned on the official GSC v2 training split (35 words) under three few-shot conditions: 15-shot, 5-shot, and 3-shot. For each condition, we randomly sampled 15/5/3 examples per keyword and repeated the sampling five times; we report the mean and variance across these subsets. Validation followed the official GSC v2 validation split.

\noindent \textbf{Test sets.} For clean testing, we used the official GSC v2 test set (10 keywords). For mixed-speech evaluation, we constructed 2-mix and 3-mix versions by randomly mixing two or three clean-test utterances with distinct keywords, using energy ratios of 1:1 and 1:1:1, respectively. The 3-mix set was used to assess generalization beyond the training mixture count.

\subsection{Models}
\label{ssec:setting}

\textbf{Pre-training configuration.} MT-HuBERT was trained from scratch. Following Cocktail-HuBERT and HuBERT-Large, the k-means codebook was derived from the 9th Transformer layer features extracted by the released \texttt{HuBERT\_BASE}\footnote{\label{hubert_base}https://dl.fbaipublicfiles.com/hubert/hubert\_base\_ls960.pt} model. We used \texttt{fairseq}\footnote{https://github.com/facebookresearch/fairseq} for training with 1.6M steps, learning rate 1e-4, 32k warmup steps, and a maximum of 700k tokens per GPU. Model size and architectural hyperparameters followed HuBERT-BASE.

\noindent \textbf{Baselines.} We compared against five representative models: HuBERT\_BASE, HuBERT-iter3, WavLM\_BASE, MPC-HuBERT, and Cocktail-HuBERT. HuBERT\_BASE and WavLM\_BASE\footnote{https://github.com/microsoft/unilm/tree/master/wavlm} were taken from their official BASE releases. HuBERT-iter3 was trained on clean speech using the same pseudo-label generation pipeline as MT-HuBERT (third HuBERT iteration on clean data). MPC-HuBERT was pre-trained on our 2-mix LibriSpeech-960h (two utterances mixed with the energy ratio of 1:1). Cocktail-HuBERT and MPC-HuBERT were trained under the same hyperparameter configuration as MT-HuBERT unless otherwise noted.

\noindent \textbf{Adaptation.} For few-shot KWS, the pre-trained backbone was frozen, and two linear layers were added and fine-tuned. We evaluated three strategies (all with BCE loss, following~\cite{yuan2024few}): \emph{Clean} (train on clean only), \emph{Mixup}~\cite{zhang2018mixup} (linearly interpolated waveforms and labels to form 2-mix), and \emph{MT} (train on clean + 2-mix with MT). All models were trained for 50 epochs using Adam with an initial LR of 0.001. The final model is the average of the last 10 checkpoints.

\noindent \textbf{Metrics.} We report Equal Error Rate (EER) for presence detection and Top-$k$ Accuracy for keyword discrimination. Consistent with mixture cardinality, we report Top-1 on clean, Top-2 on 2-mix, and Top-3 on 3-mix.

\noindent \noindent \textbf{Code.} Our implementation and scripts to reproduce all experiments are available at \url{https://github.com/asip-cslt/MT-HuBERT}.

\section{Experimental Results}
\label{sec:result}

\begin{table*}[htp]
  \centering
  \vspace{-3mm}
  \caption{Few-shot KWS performance on clean, 2-mix and 3-mix tests}
  \label{tab:tab_1}
  \scalebox{0.84}{
    \begin{tabular}{cccccccccc}
    \toprule
    \multicolumn{10}{c}{\textbf{(a) Top-1 ACC(\%) and EER(\%) with clean test}} \\
    \midrule
    \multirow{2}{*}{Pre-train} & \multicolumn{3}{c}{Adaptation} & \multicolumn{2}{c}{15-shot} & \multicolumn{2}{c}{5-shot} & \multicolumn{2}{c}{3-shot}   \\
    \cmidrule(r){2-4} \cmidrule(r){5-6} \cmidrule(r){7-8} \cmidrule(r){9-10}
                & Clean & Mixup & MT & ACC ($\uparrow$) & EER ($\downarrow$) & ACC ($\uparrow$) & EER ($\downarrow$) & ACC ($\uparrow$) & EER ($\downarrow$) \\
    \cmidrule(r){1-1} \cmidrule(r){2-4}	\cmidrule(r){5-6}   \cmidrule(r){7-8}  \cmidrule(r){9-10}
                \multirow{3}{*}{HuBERT\_BASE~\cite{hsu2021hubert}} & \checkmark &&&  89.09$\pm$0.35  & 4.47$\pm$0.05 & 75.26$\pm$2.56  & 8.70$\pm$0.25  & 58.39$\pm$7.63  & 14.55$\pm$0.96 \\
                & & \checkmark & &                 90.17$\pm$0.57  & 4.08$\pm$0.04 & 77.71$\pm$2.22  & 7.61$\pm$0.18  & 60.96$\pm$7.95  & 13.12$\pm$0.71 \\
                & & & \checkmark &                 91.13$\pm$0.24  & 3.92$\pm$0.03 & 80.69$\pm$1.85  & 7.10$\pm$0.22 & 65.30$\pm$10.20 & 11.11$\pm$0.83 \\
    \cmidrule(r){1-1}   \cmidrule(r){2-4}	 \cmidrule(r){5-6}   \cmidrule(r){7-8}  \cmidrule(r){9-10}
                \multirow{3}{*}{HuBERT-iter3~\cite{hsu2021hubert}} & \checkmark &&&  91.44$\pm$0.27  & 3.51$\pm$0.03 & 87.60$\pm$1.28  & 5.60$\pm$0.09  & 80.17$\pm$4.68  & 8.25$\pm$1.25 \\
                & & \checkmark & &                 91.94$\pm$0.03  & 3.14$\pm$0.03 & 88.36$\pm$2.18  & 4.88$\pm$0.41  & 82.39$\pm$5.95  & 6.98$\pm$1.22 \\
                & & & \checkmark &                 92.05$\pm$0.30  & 3.42$\pm$0.06 & 88.22$\pm$1.79  & 4.94$\pm$0.37 & 81.94$\pm$9.08  & 6.81$\pm$1.02 \\
    \cmidrule(r){1-1}   \cmidrule(r){2-4}	 \cmidrule(r){5-6}   \cmidrule(r){7-8}  \cmidrule(r){9-10}
                \multirow{3}{*}{WavLM\_BASE~\cite{chen2022wavlm}} & \checkmark &&&  91.28$\pm$0.07  & 4.10$\pm$0.01 & 80.82$\pm$2.13  & 7.43$\pm$0.16  & 66.09$\pm$6.27  & 12.12$\pm$1.63 \\
                & & \checkmark & &                 92.46$\pm$0.09  & 3.59$\pm$0.06 & 83.79$\pm$0.43  & 6.49$\pm$0.04  & 71.13$\pm$8.08  & 10.52$\pm$1.70 \\
                & & & \checkmark &                 92.84$\pm$0.07  & 3.55$\pm$0.02 & 86.00$\pm$0.86  & 5.93$\pm$0.04 & 73.28$\pm$2.24  & 9.58$\pm$1.02 \\
    \cmidrule(r){1-1}   \cmidrule(r){2-4}	 \cmidrule(r){5-6}   \cmidrule(r){7-8}  \cmidrule(r){9-10}
                \multirow{3}{*}{MPC-HuBERT~\cite{wang2024speech}} & \checkmark &&&  89.14$\pm$0.48  & 4.48$\pm$0.17 & 79.68$\pm$6.52  & 7.99$\pm$0.51  & 68.81$\pm$8.42  & 11.19$\pm$1.95 \\
                & & \checkmark & &                 89.99$\pm$0.19  & 3.98$\pm$0.10 & 81.78$\pm$4.31  & 6.69$\pm$0.27  & 72.72$\pm$0.84  & 9.31$\pm$0.17 \\
                & & & \checkmark &                 90.81$\pm$0.36  & 3.84$\pm$0.12 & 84.87$\pm$2.11  & 5.99$\pm$0.10 & 76.03$\pm$2.38  & 8.10$\pm$0.47 \\
    \cmidrule(r){1-1}   \cmidrule(r){2-4}	 \cmidrule(r){5-6}   \cmidrule(r){7-8}  \cmidrule(r){9-10}
                \multirow{3}{*}{Cocktail-HuBERT~\cite{fazel2023cocktail}} & \checkmark &&&                                88.04$\pm$0.21  & 5.05$\pm$0.03 & 77.97$\pm$2.58  & 8.75$\pm$0.62  & 65.54$\pm$4.80  & 12.44$\pm$0.76 \\
                & & \checkmark & &                 89.96$\pm$0.63  & 4.15$\pm$0.06 & 82.32$\pm$5.33  & 6.76$\pm$0.51  & 71.51$\pm$2.71  & 10.11$\pm$0.52 \\
                & & & \checkmark &                 90.79$\pm$0.25  & 4.09$\pm$0.06 & 84.10$\pm$8.30  & 6.16$\pm$0.64 & 73.53$\pm$0.96  & 9.41$\pm$0.12 \\
    \cmidrule(r){1-1}   \cmidrule(r){2-4}	 \cmidrule(r){5-6}   \cmidrule(r){7-8}  \cmidrule(r){9-10}
                \multirow{3}{*}{MT-HuBERT (Ours)} & \checkmark &&&  92.60$\pm$0.15  & 3.51$\pm$0.02 & 88.68$\pm$2.84  & 5.30$\pm$0.24  & 79.95$\pm$6.47  & 7.89$\pm$0.64 \\
                & & \checkmark & &                 92.89$\pm$0.21  & 3.15$\pm$0.03 & 90.15$\pm$0.87  & 4.43$\pm$0.12  & 82.19$\pm$4.00  & 6.99$\pm$0.32 \\
                & & & \checkmark &                 \textbf{93.80$\pm$0.04}  & \textbf{2.95$\pm$0.01} & \textbf{91.55$\pm$0.74}  & \textbf{4.01$\pm$0.11} & \textbf{83.58$\pm$3.72}  & \textbf{6.41$\pm$0.55} \\
    \midrule       
    \midrule       
    \multicolumn{10}{c}{\textbf{(b) Top-2 ACC(\%) and EER(\%) with 2-mix test}} \\
    \midrule
    \multirow{2}{*}{Pre-train} & \multicolumn{3}{c}{Adaptation} & \multicolumn{2}{c}{15-shot} & \multicolumn{2}{c}{5-shot} & \multicolumn{2}{c}{3-shot}   \\
    \cmidrule(r){2-4} \cmidrule(r){5-6} \cmidrule(r){7-8} \cmidrule(r){9-10}
                & Clean & Mixup & MT & ACC ($\uparrow$) & EER ($\downarrow$) & ACC ($\uparrow$) & EER ($\downarrow$) & ACC ($\uparrow$) & EER ($\downarrow$) \\
    \cmidrule(r){1-1} \cmidrule(r){2-4}	\cmidrule(r){5-6}   \cmidrule(r){7-8}  \cmidrule(r){9-10}
                \multirow{3}{*}{HuBERT\_BASE~\cite{hsu2021hubert}} & \checkmark &&&  48.64$\pm$1.49  & 21.29$\pm$0.72  & 40.45$\pm$3.25  & 25.37$\pm$1.36  & 29.37$\pm$13.12 & 32.24$\pm$6.08 \\
                & & \checkmark & &                 57.04$\pm$0.99  & 17.09$\pm$0.26  & 48.54$\pm$1.62  & 20.60$\pm$0.23  & 34.34$\pm$13.94 & 28.64$\pm$1.42 \\
                & & & \checkmark &                61.74$\pm$0.60  & 15.31$\pm$0.14  & 51.86$\pm$5.08  & 19.16$\pm$0.69  & 41.25$\pm$6.70  & 23.92$\pm$1.01 \\
    \cmidrule(r){1-1} \cmidrule(r){2-4}	\cmidrule(r){5-6}   \cmidrule(r){7-8}  \cmidrule(r){9-10}
                \multirow{3}{*}{HuBERT-iter3~\cite{hsu2021hubert}} & \checkmark &&&  55.79$\pm$0.21  & 19.28$\pm$0.08  & 52.07$\pm$2.29  & 21.16$\pm$0.45  & 45.37$\pm$7.13  & 24.57$\pm$1.68 \\
                & & \checkmark & &                 62.67$\pm$0.27  & 15.95$\pm$0.05  & 57.95$\pm$1.60  & 17.82$\pm$0.23  & 52.39$\pm$1.39  & 20.22$\pm$0.44 \\
                & & & \checkmark &                65.37$\pm$0.64  & 14.62$\pm$0.11  & 59.61$\pm$3.13  & 17.15$\pm$0.65  & 54.84$\pm$0.34  & 19.07$\pm$0.20 \\
    \cmidrule(r){1-1} \cmidrule(r){2-4}	\cmidrule(r){5-6}   \cmidrule(r){7-8}  \cmidrule(r){9-10}
                \multirow{3}{*}{WavLM\_BASE~\cite{chen2022wavlm}} & \checkmark &&&  54.79$\pm$0.85  & 22.53$\pm$0.30  & 49.78$\pm$1.58  & 24.12$\pm$0.29  & 42.63$\pm$2.72  & 27.61$\pm$1.26 \\
                & & \checkmark & &                 58.17$\pm$0.22  & 21.02$\pm$0.04  & 52.56$\pm$0.30  & 23.02$\pm$0.15  & 46.58$\pm$1.45  & 25.41$\pm$0.81 \\
                & & & \checkmark &                59.71$\pm$0.07  & 20.36$\pm$0.07  & 54.94$\pm$0.57  & 22.30$\pm$0.15  & 48.31$\pm$0.47  & 24.39$\pm$0.43 \\
    \cmidrule(r){1-1} \cmidrule(r){2-4}	\cmidrule(r){5-6}   \cmidrule(r){7-8}  \cmidrule(r){9-10}
                \multirow{3}{*}{MPC-HuBERT~\cite{wang2024speech}} & \checkmark &&&  58.00$\pm$1.58  & 18.41$\pm$0.49  & 53.71$\pm$5.24  & 20.00$\pm$1.53  & 42.03$\pm$21.04 & 25.99$\pm$6.75 \\
                & & \checkmark & &                 66.33$\pm$0.58  & 14.07$\pm$0.14  & 60.08$\pm$2.29  & 16.32$\pm$0.28  & 49.24$\pm$6.56  & 20.88$\pm$0.69 \\
                & & & \checkmark &                71.26$\pm$1.45  & 12.26$\pm$0.24  & 63.95$\pm$1.76  & 14.95$\pm$0.06  & 54.22$\pm$5.65  & 18.46$\pm$0.61 \\
    \cmidrule(r){1-1} \cmidrule(r){2-4}	\cmidrule(r){5-6}   \cmidrule(r){7-8}  \cmidrule(r){9-10}
                \multirow{3}{*}{Cocktail-HuBERT~\cite{fazel2023cocktail}} & \checkmark &&&                               68.10$\pm$1.13  & 13.25$\pm$0.43  & 59.73$\pm$3.55  & 16.22$\pm$1.49  & 48.87$\pm$2.17  & 21.23$\pm$1.79 \\
                & & \checkmark & &                 74.30$\pm$1.45  & 10.79$\pm$0.22  & 68.51$\pm$2.92  & 12.54$\pm$0.32  & 60.08$\pm$3.93  & 15.95$\pm$0.55 \\
                & & & \checkmark &                78.54$\pm$0.62  & 9.29$\pm$0.16  & 71.27$\pm$5.96  & 11.88$\pm$0.72  & 62.34$\pm$2.43  & 15.30$\pm$0.23 \\
    \cmidrule(r){1-1} \cmidrule(r){2-4}	\cmidrule(r){5-6}   \cmidrule(r){7-8}  \cmidrule(r){9-10}
                \multirow{3}{*}{MT-HuBERT (Ours)} & \checkmark &&&  71.49$\pm$0.60  & 12.68$\pm$0.14  & 65.99$\pm$4.00  & 14.24$\pm$0.79  & 57.60$\pm$2.00  & 17.68$\pm$0.83 \\
                & & \checkmark & &                 76.76$\pm$1.16  & 10.23$\pm$0.37  & 71.81$\pm$0.77  & 11.89$\pm$0.13  & 63.67$\pm$2.06  & 14.86$\pm$0.51 \\
                & & & \checkmark &                \textbf{79.78$\pm$0.38}  & \textbf{8.98$\pm$0.13}  & \textbf{75.00$\pm$2.36}  & \textbf{11.15$\pm$0.48}  & \textbf{66.63$\pm$0.83}  & \textbf{13.95$\pm$0.17} \\
    \midrule       
    \midrule       
    \multicolumn{10}{c}{\textbf{(c) Top-3 ACC(\%) and EER(\%) with 3-mix test}} \\
    \midrule
    \multirow{2}{*}{Pre-train} & \multicolumn{3}{c}{Adaptation} & \multicolumn{2}{c}{15-shot} & \multicolumn{2}{c}{5-shot} & \multicolumn{2}{c}{3-shot}   \\
    \cmidrule(r){2-4} \cmidrule(r){5-6} \cmidrule(r){7-8} \cmidrule(r){9-10}
                & Clean & Mixup & MT & ACC ($\uparrow$) & EER ($\downarrow$) & ACC ($\uparrow$) & EER ($\downarrow$) & ACC ($\uparrow$) & EER ($\downarrow$) \\
    \cmidrule(r){1-1} \cmidrule(r){2-4}	\cmidrule(r){5-6}   \cmidrule(r){7-8}  \cmidrule(r){9-10}
                \multirow{3}{*}{HuBERT\_BASE~\cite{hsu2021hubert}} & \checkmark &&&  33.19$\pm$2.73  & 32.22$\pm$1.82 & 26.22$\pm$3.86  & 36.42$\pm$1.86  & 18.40$\pm$11.17 & 42.31$\pm$9.48 \\
                & & \checkmark & &                 43.09$\pm$1.09  & 26.48$\pm$0.95 & 35.97$\pm$1.46  & 30.16$\pm$0.39  & 24.17$\pm$6.52  & 37.09$\pm$1.72 \\
                & & & \checkmark &                 47.15$\pm$0.39  & 24.60$\pm$0.30 & 39.29$\pm$1.79  & 28.23$\pm$0.26 & 29.68$\pm$6.40  & 33.04$\pm$2.25 \\
    \cmidrule(r){1-1}   \cmidrule(r){2-4}	 \cmidrule(r){5-6}   \cmidrule(r){7-8}  \cmidrule(r){9-10}
                \multirow{3}{*}{HuBERT-iter3~\cite{hsu2021hubert}} & \checkmark &&&  40.76$\pm$1.45  & 29.14$\pm$0.24 & 37.89$\pm$5.03  & 30.27$\pm$1.85  & 32.43$\pm$9.49  & 33.64$\pm$4.31 \\
                & & \checkmark & &                 47.11$\pm$0.26  & 26.36$\pm$0.32 & 45.10$\pm$3.46  & 26.37$\pm$1.45  & 40.59$\pm$1.01  & 28.81$\pm$0.65 \\
                & & & \checkmark &                 49.87$\pm$0.22  & 24.57$\pm$0.23 & 46.17$\pm$4.24  & 26.03$\pm$1.37 & 41.76$\pm$0.33  & 28.03$\pm$0.61 \\
    \cmidrule(r){1-1}   \cmidrule(r){2-4}	 \cmidrule(r){5-6}   \cmidrule(r){7-8}  \cmidrule(r){9-10}
                \multirow{3}{*}{WavLM\_BASE~\cite{chen2022wavlm}} & \checkmark &&&  43.24$\pm$1.21  & 29.63$\pm$0.82 & 39.99$\pm$1.87  & 30.85$\pm$1.19 & 34.76$\pm$4.08  & 34.24$\pm$2.23 \\
                & & \checkmark & &                 46.14$\pm$0.61  & 28.71$\pm$0.57 & 42.44$\pm$0.38  & 29.87$\pm$0.07  & 38.17$\pm$4.29  & 31.90$\pm$2.28 \\
                & & & \checkmark &                 47.34$\pm$0.29  & 28.14$\pm$0.24 & 43.89$\pm$1.27  & 29.30$\pm$0.54 & 39.09$\pm$1.44  & 31.06$\pm$1.25 \\
    \cmidrule(r){1-1}   \cmidrule(r){2-4}	 \cmidrule(r){5-6}   \cmidrule(r){7-8}  \cmidrule(r){9-10}
                \multirow{3}{*}{MPC-HuBERT~\cite{wang2024speech}} & \checkmark &&&  43.90$\pm$3.78  & 27.79$\pm$0.34 & 42.96$\pm$3.75  & 28.10$\pm$2.44  & 31.55$\pm$19.92 & 34.40$\pm$6.88 \\
                & & \checkmark & &                 54.23$\pm$0.46  & 21.45$\pm$0.35 & 49.72$\pm$2.89  & 23.27$\pm$0.56  & 39.13$\pm$9.83  & 28.81$\pm$3.37 \\
                & & & \checkmark &                 56.58$\pm$0.05  & 20.53$\pm$0.33 & 51.85$\pm$2.97  & 22.23$\pm$0.58 & 43.67$\pm$3.50  & 25.93$\pm$1.17 \\
    \cmidrule(r){1-1}   \cmidrule(r){2-4}	 \cmidrule(r){5-6}   \cmidrule(r){7-8}  \cmidrule(r){9-10}
                \multirow{3}{*}{Cocktail-HuBERT~\cite{fazel2023cocktail}} & \checkmark &&&                                55.12$\pm$0.71  & 20.26$\pm$0.42 & 50.39$\pm$2.08  & 22.14$\pm$2.90  & 40.65$\pm$6.38  & 27.87$\pm$2.46 \\
                & & \checkmark & &                 59.57$\pm$0.65  & 18.55$\pm$0.21 & 56.46$\pm$4.86  & 19.61$\pm$1.40  & 49.76$\pm$2.51  & 22.79$\pm$0.38 \\
                & & & \checkmark &                 62.65$\pm$0.65  & 17.31$\pm$0.27 & 59.37$\pm$2.52  & 18.21$\pm$0.53 & 51.78$\pm$1.18  & 21.66$\pm$0.27 \\
    \cmidrule(r){1-1}   \cmidrule(r){2-4}	 \cmidrule(r){5-6}   \cmidrule(r){7-8}  \cmidrule(r){9-10}
                \multirow{3}{*}{MT-HuBERT (Ours)} & \checkmark &&&  57.90$\pm$2.31  & 19.12$\pm$0.77 & 53.42$\pm$6.04  & 21.12$\pm$1.89  & 46.60$\pm$2.48  & 24.40$\pm$1.25 \\
                & & \checkmark & &                 63.28$\pm$1.95  & 17.04$\pm$0.76 & 60.25$\pm$2.94  & 17.81$\pm$0.74  & 53.55$\pm$1.93  & 20.78$\pm$1.01 \\
                & & & \checkmark &                 \textbf{65.91$\pm$0.73}  & \textbf{15.99$\pm$0.40} & \textbf{62.00$\pm$3.47}  & \textbf{17.23$\pm$1.15} & \textbf{54.95$\pm$2.20}  & \textbf{20.05$\pm$1.03} \\
    \bottomrule
    \end{tabular}}
\end{table*}

Table~\ref{tab:tab_1} reports few-shot KWS under clean (Top-1), 2-mix (Top-2), and 3-mix (Top-3) tests across different pre-training backbones and adaptation strategies (Clean / Mixup / MT). We summarize four consistent findings.

\noindent \textbf{1. MT-HuBERT is the best pre-training model.}
First, fixing MT as the adaptation, MT-HuBERT is consistently the strongest, especially under 2-mix and unseen 3-mix cardinality. For 2-mix (15-shot), the ranking is MT-HuBERT $>$ Cocktail-HuBERT $>$ MPC-HuBERT $>$ HuBERT-it3 $>$ HuBERT\_BASE $>$ WavLM\_BASE. The same qualitative ordering holds under Mixup or Clean adaptation, with uniformly worse absolute performance than MT.

\noindent \textbf{2. MT is the best adaptation approach.}
Then, fixing a pre-trained backbone, MT adaptation $>$ Mixup $>$ Clean across tests, with larger gains under 2-/3-mix. For MT-HuBERT (15-shot), Clean $\rightarrow$ Mixup $\rightarrow$ MT yields clean: 92.60/3.51 $\rightarrow$ 92.89/3.15 $\rightarrow$ 93.80/2.95. Trends persist for 5-shot and 3-shot and for other backbones.

\noindent \textbf{3. MT-HuBERT pre-training + MT adaptation yields the best overall results.} The best overall combination is MT-HuBERT (pre-train) + MT (adaptation). It attains 93.80/2.95 on clean, 79.78/8.98 on 2-mix, and 65.91/15.99 on 3-mix (15-shot), exceeding its strongest competitor Cocktail-HuBERT+MT by +1.24/-0.31 (ACC/EER) on 2-mix and +3.26/-1.32 on 3-mix; the 2-mix margins further widen at 5-shot (+3.73/-0.73) and 3-shot (+4.29/-1.35).

\noindent \textbf{4. Benefits increase in more challenging scenarios.}
None of the models saw 3-mix during pre-training or adaptation. Most clean-only (HuBERT series) or denoising-style (WavLM\_BASE) pre-training degrades below 50\% Top-3 ACC in 3-mix, whereas MT-HuBERT + MT achieves 65.91 (15-shot), 62.00 (5-shot), and 54.95 (3-shot), consistently ahead of Cocktail-HuBERT+MT (62.65, 59.37, 51.78). These results support our claim that $k$-hot masked prediction against a clean-speech codebook encourages the backbone to represent mixed speech by \emph{composition of clean acoustic units}, rather than single mixed patterns, yielding stronger generalization to unseen mixing scenarios.

In summary, integrating SSL pre-training with MT is effective for mixed-speech KWS: MT-HuBERT + MT consistently delivers the highest ACC and lowest EER across clean/2-mix/3-mix and across shots. Its advantage is more pronounced in lower-shot regimes and under heavier overlap, further demonstrating its effectiveness. These observations are consistent with prior findings~\cite{yuan2024few}, indicating that MT is a general solution for tackling mixed-speech processing in both supervised and self-supervised modes.

\section{Conclusion}
\label{sec:conclusion}

This paper addressed few-shot keyword spotting in mixed speech under a pre-training and adaptation paradigm. We introduced MT-HuBERT, a self-supervised pre-training framework that implements the Mix-Training (MT) criterion at pre-training time via $k$-hot masked prediction against a clean-speech codebook, encouraging the backbone to represent mixed speech by composition of clean acoustic units rather than mixed patterns. Coupled with MT adaptation, MT-HuBERT achieves state-of-the-art results among the compared baselines on GSC v2 across clean (Top-1), 2-mix (Top-2), and unseen 3-mix (Top-3) evaluations, with gains most pronounced in low-shot and higher-overlap regimes, while remaining competitive on clean speech. Promising directions include scaling mixture cardinality and SNR diversity during pre-training and testing, extending to larger and multilingual corpora, and evaluating transfer to other speech tasks (e.g., ASR).

\vfill\pagebreak

\bibliographystyle{IEEEbib}
\bibliography{ref}

\end{document}